\documentclass[aps,prl,twocolumn,amsmath,amsart,amssymb,superscriptaddress]{revtex4-1}

\usepackage{graphicx}
\usepackage{latexsym}
\usepackage[T1]{fontenc}
\usepackage{amsmath}
\usepackage{endnotes}
\usepackage{lmodern}
\usepackage{color}

\newcommand{\SIOS}{Sr$_2$IrO$_4$}
\newcommand{\SION}{Sr$_3$Ir$_2$O$_7$}

\newcommand{\jeff}{$J_{\textrm{eff}} = \frac{1}{2}$ }
\newcommand{\jeffh}{$j = \frac{1}{2}$ }
\newcommand{\jeffth}{$j = \frac{3}{2}$ }
\begin{document}

\title{Dispersive Magnetic and Electronic Excitations in Iridate Perovskites Probed with Oxygen $K$-Edge Resonant Inelastic X-ray Scattering}
\author{Xingye Lu}
\email{xingye.lu@psi.ch}
\thanks{\\Present address: Department of Physics, Beijing Normal University, Beijing 100875, China.}
\affiliation{Research Department Synchrotron Radiation and Nanotechnology, Paul Scherrer Institut, CH-5232 Villigen PSI, Switzerland}

\author{Paul Olalde-Velasco}
\thanks{Present address: Instituto de F{\' i}sica, Benem{\'e}rita Universidad Aut{\'o}noma de Puebla, Apdo. Postal J-48, Puebla, Puebla 72570, M{\'e}xico.}
\affiliation{Research Department Synchrotron Radiation and Nanotechnology, Paul Scherrer Institut, CH-5232 Villigen PSI, Switzerland}

\author{Yaobo Huang}
\affiliation{Research Department Synchrotron Radiation and Nanotechnology, Paul Scherrer Institut, CH-5232 Villigen PSI, Switzerland}

\author{Valentina Bisogni}
\thanks{Present address: National Synchrotron Light Source II, Brookhaven National Laboratory, Upton, New York 11973, USA.}
\affiliation{Research Department Synchrotron Radiation and Nanotechnology, Paul Scherrer Institut, CH-5232 Villigen PSI, Switzerland}

\author{Jonathan Pelliciari}
\thanks{Present address: Department of Physics, Massachusetts Institute of Technology, Cambridge, Massachusetts 02139, USA}
\affiliation{Research Department Synchrotron Radiation and Nanotechnology, Paul Scherrer Institut, CH-5232 Villigen PSI, Switzerland}

\author{Sara Fatale}

\affiliation{Laboratory for Quantum Magnetism, Institute of Physics, Ecole Polytechnique F\'{e}derale de Lausanne (EPFL), CH-1015 Lausanne, Switzerland}

\author{Marcus Dantz}
\affiliation{Research Department Synchrotron Radiation and Nanotechnology, Paul Scherrer Institut, CH-5232 Villigen PSI, Switzerland}

\author{James G. Vale}
\affiliation{London Centre for Nanotechnology and Department of Physics and Astronomy, University College London, Gower Street, London, WC1E 6BT, United Kingdom}

\author{E. C. Hunter}
\affiliation{School of Physics and Astronomy, The University of Edinburgh, James Clerk Maxwell Building, Mayfield Road, Edinburgh EH9 2TT, United Kingdom}

\author{Johan Chang}
\affiliation{Laboratory for Quantum Magnetism, Institute of Condensed Matter Physics (ICMP), Ecole Polytechnique F\'{e}derale de Lausanne (EPFL), CH-1015 Lausanne, Switzerland}

\author{Vladimir N. Strocov}
\affiliation{Research Department Synchrotron Radiation and Nanotechnology, Paul Scherrer Institut, CH-5232 Villigen PSI, Switzerland}

\author{R. S. Perry}
\affiliation{ISIS neutron spallation source, Rutherford Appleton Laboratory (RAL), Harwell Campus, Didcot OX11 0QX, United Kingdom}
\affiliation{London Centre for Nanotechnology and UCL Centre for Materials Discovery, University College London, 17-19 Gordon Street, London WC1H 0AH, United Kingdom}

\author{Marco Grioni}
\affiliation{Laboratory for Quantum Magnetism, Institute of Condensed Matter Physics (ICMP), Ecole Polytechnique F\'{e}derale de Lausanne (EPFL), CH-1015 Lausanne, Switzerland}

\author{D. F.  McMorrow}
\affiliation{London Centre for Nanotechnology and Department of Physics and Astronomy, University College London, Gower Street, London, WC1E 6BT, United Kingdom}

\author{Henrik M. R{\o}nnow}
\affiliation{Laboratory for Quantum Magnetism, Institute of Physics (ICMP), Ecole Polytechnique F\'{e}derale de Lausanne (EPFL), CH-1015 Lausanne, Switzerland}

\author{Thorsten Schmitt}
\email{thorsten.schmitt@psi.ch}
\affiliation{Research Department Synchrotron Radiation and Nanotechnology, Paul Scherrer Institut, CH-5232 Villigen PSI, Switzerland}

\date{\today}

\begin{abstract}
Resonant inelastic X-ray scattering (RIXS) experiments performed at the oxygen-$K$ edge on the iridate perovskites {\SIOS} and {\SION} 
reveal a sequence of well-defined dispersive modes over the energy range up to $\sim 0.8$ eV. 
The momentum dependence of these modes and their variation with the experimental geometry allows us to assign each of them to specific collective magnetic and/or electronic excitation processes, including single and bi-magnons, and spin-orbit and electron-hole excitons. We thus demonstrated that dispersive magnetic and electronic excitations are observable at the 
O-$K$ edge in the presence of the strong spin-orbit coupling in the $5d$ shell of iridium and strong hybridization between Ir $5d$ and O $2p$ orbitals, which confirm and expand theoretical expectations. More generally, our results establish the utility of O-$K$ edge RIXS for studying the collective excitations in a range of $5d$ materials
that are attracting increasing attention due to their novel magnetic and electronic properties. Especially, the strong RIXS response at O-$K$ edge opens up the opportunity for investigating collective excitations in thin films and heterostructures fabricated from these materials.
\end{abstract}

\maketitle

Characterizing the elementary excitations in correlated electron systems is an essential prerequisite for obtaining a complete understanding of the underlying electronic interactions and therefore crucial for revealing the origin of emergent phases \cite{anderson, rmp_rixs}.  Resonant inelastic X-ray scattering (RIXS) has become established in the past decade as a powerful tool for studying the momentum dependence of electronic and magnetic excitations \cite{rmp_rixs}. Of particular interest is RIXS at the transition-metal (TM) $L_{2, 3}$ edges. Because of the strong spin-orbit coupling (SOC) in the $2p_{\frac{3}{2}, \frac{1}{2}}$ level of the intermediate state, single spin-flip excitations are directly allowed at TM-$L_{2, 3}$ edges \cite{ament_09, LB_09, LB_10}, making TM-$L_{2, 3}$ RIXS specially suited for investigating the magnetic dynamics in 3$d$ (cuprates and iron-pnictides) \cite{ament_09, LB_09, LB_10, tacon_11, nature_12, dean_12, dean_13, kejin_13, weisheng_14, chunjing_ncom, grioni_14, monney_16} and 5$d$ (iridates and osmates) systems \cite{jkim_214,jkim_327,bjkim_14ncom,marco_327,calder}.

While the Mott state in cuprates is dominated by strong on-site electron correlation ($U$), the \jeff Mott state in $5d^5$ iridates is driven by collaborative strong SOC ($\sim 0.5$ eV) and intermediate $U (\sim 2$ eV) under strong octahedral crystal electric field \cite{bjkim_08}. Because of their novel Mott physics and certain parallelism with cuprates, layered iridates ({\SIOS} and {\SION}) have attracted substantial research interest \cite{bjkim_08, jkim_214, jkim_327, bjkim_14ncom, marco_327} [Figs. 1(a)-(c)]. More recently, thin films and heterostructures of perovskite iridates have been suggested to host non-trivial topological states \cite{xiaodi,hykee_ncom,fang_nphys} and triggered a new wave of studies on artificial structures \cite{takagi_113, yjkim, seo}. However, measuring elementary excitations using Ir-$L$ RIXS on artificial low-dimensional samples requires longer counting time than that on bulk crystal because $11$ keV of X-ray has $\sim 10\mu$m \cite{length} of attenuation length while artificial thin film and superlattice are usually much thinner ($\sim 10-100$nm) \cite{dean_12, dean_13, yjkim, seo}. 

\begin{figure}[htbp]
\includegraphics[width=8.5cm]{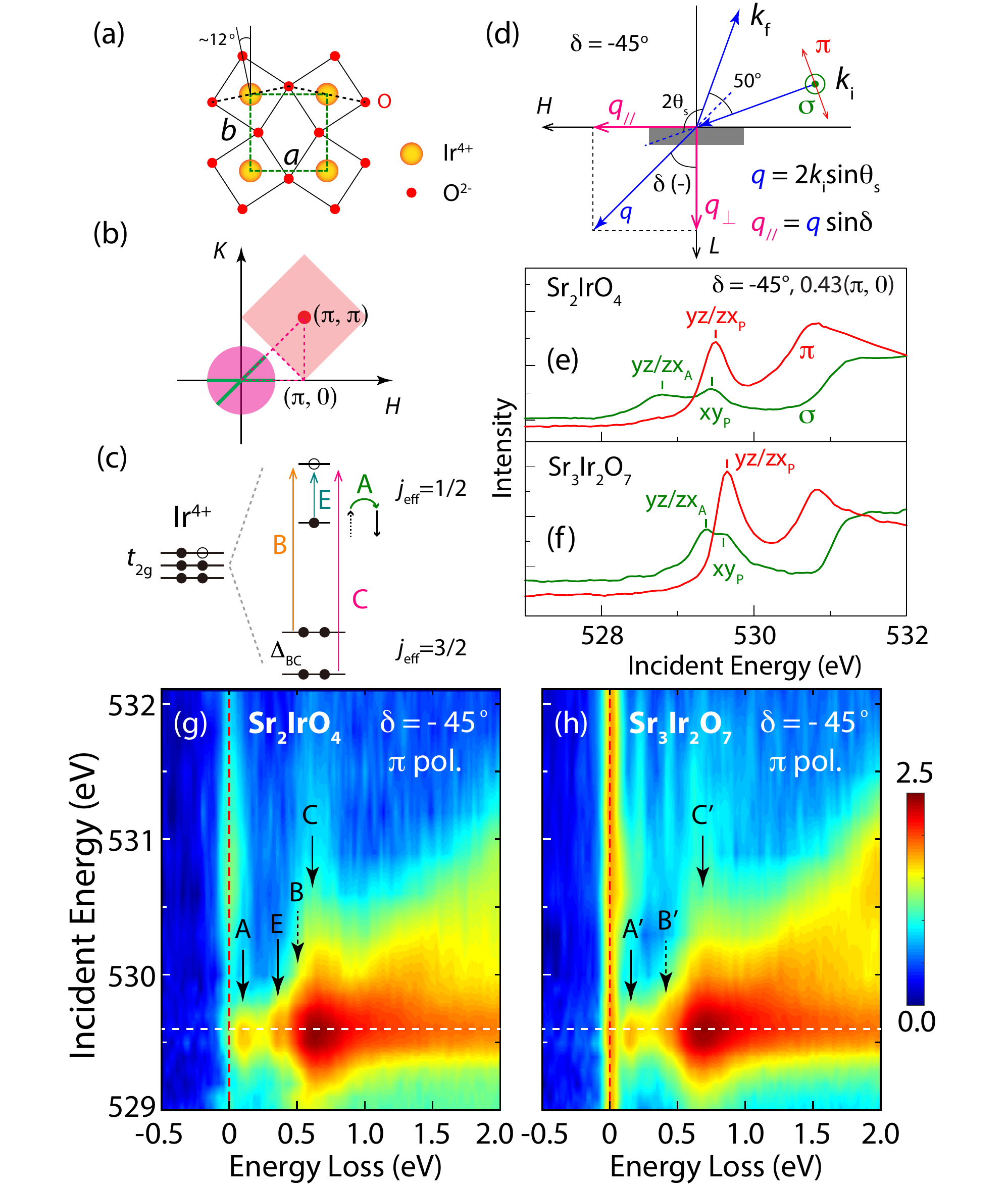}
\caption{(color online). (a) Schematic of the distorted IrO plane of {\SIOS} and {\SION}. (b) The corresponding reciprocal space in tetragonal notation. The pink filled circle around $\Gamma$ point is the momentum coverage of O-$K$ RIXS with 2$\theta_s$ = 130$^{\text{o}}$. The green solid lines mark the two directions [H, 0] and [H, H] we measured. (c) Schematic of elementary excitations of iridates in a single-ion picture.  A denotes single magnon, B and C spin-orbit excitons and E electron-hole exciton. $\Delta_{\text{BC}}$ marks the energy difference between B and C.  (d) Schematics of back-scattering RIXS experimental setup. The blue arrows $k_i$ and $k_f$ denote the incident and scattered photons, where the polarizations are shown as red arrow ($\pi $) and green circle-dot ($\sigma$). The scattering angle is 2$\theta_s$ = 130$^{\text{o}}$. The projections of ${\it q}$ onto two reciprocal axes ($q_{\parallel}$ and $q_{\perp}$) can be tuned continuously by changing $\delta$.  (e)-(f) X-ray absorption spectra at a grazing-incidence angle ($\delta = -45^{\text{o}}$) along [H, 0]. The probed Ir $t_{2g}$ orbitals (xy, yz and zx) are marked, where the subscripts A and P indicates that orbitals are probed via their hybridization with apical or plane oxygens. (g)-(h) Incident-energy dependence of the elementary excitations of {\SIOS} and {\SION}, measured near the O-$K$ edge with $\pi$ polarization and $\delta = -45^{\text{o}}$ along [H, 0].}
\label{fig.F1}
\end{figure}

RIXS at O-$K$ edge (O $1s$ to O $2p$, $\sim 530$ eV) offers unique opportunities in probing the elementary excitations in $4d$ and $5d$ TM oxides \cite{marco_14, chang_15} with high energy resolution ( $\sim$ 45 meV) \cite{weisheng_13, note}. Although O-$K$ RIXS has much smaller attenuation length ($\lesssim100$nm) than Ir-$L$ RIXS, a comparison between O-$K$ and Ir-$L_3$ spectra measured on {\SIOS} suggests that they have comparable counting efficiency (SFig. 6 in \cite{SI}). Therefore, soft x-ray RIXS is a promising method for studying artificial two-dimensional samples \cite{dean_12, dean_13}. However, given that SOC is absent in O ${1s}$ level, O-$K$ RIXS is expected to be insensitive to single spin-flip excitations \cite{valentina_OK_1, valentina_OK_2, chang_15}. This would severely limit the applicability of O-$K$ RIXS on iridates and related 5$d$ TM oxides, especially thin films and heterostructures fabricated by these materials, where magnetic excitations need to be studied for probing the low energy Hamiltonian \cite{jkim_214,jkim_327,marco_327,bjkim_14ncom}. 
 
Although SOC is absent in O $1s$ orbital, it exists in the Ir $5d$ shell which strongly hybridizes with O $2p$ orbitals. This could allow single spin-flip processes at O-$K$ edge. Indeed, Kim {\it et al.} \cite{Brink_15} proposed that single spin-flip excitations will be intense at $\Gamma$ point for O-$K$ RIXS if strong SOC is present in TM $d$ orbitals and space inversion symmetry at oxygen sites is broken. However, the symmetry analysis in Ref. \cite{Brink_15} is only viable for $q$ = 0 and $\pi$ of a one-dimensional distorted Ir$_1$-O$_1$-Ir$_2$-O$_2$ periodic arrangement [black dashed line in Fig. 1(a)], which is far from the situations for real iridate materials, such as the Perovskite layered iridates ({\SIOS} and {\SION}) bearing special interests. Later, O-$K$ RIXS measurements on a {\SIOS} thin film reported the possible observation of single magnons at low $q$ but failed to observe the corresponding dispersion that can conclusively determine the single-magnon nature of the observed signal \cite{liu_15}. Therefore, the capability of O-$K$ RIXS on probing magnetic excitations of iridates remain to be explored experimentally. In addition, the origin of some collective modes being important for understanding the charge dynamics, is still under debate \cite{bjkim_14ncom, jt_prl} and requires further experimental studies.

In this study, we use O-$K$ RIXS to study layered perovskite iridates {\SIOS} and {\SION} (the $n=1$ and $2$ of the Ruddlesden-Popper series Sr$_{n+1}$Ir$_n$O$_{3n+1}$), which exhibit novel \jeff Mott physics \cite{bjkim_08, bjkim_14ncom, jackeli_prl, review15}. We unambiguously observe the single magnon dispersion of both samples, thus providing conclusive evidence for the capability of O-$K$ RIXS in probing single magnons in 5$d$ TM oxides with large SOC. The collective excitonic quasiparticles dispersing between $0.4-0.6$ eV at the Ir-$L_3$ edge of {\SIOS} \cite{bjkim_14ncom} ``dressed'' by magnons \cite{Brink_15} have also been observed at the O-$K$ edge of {\SIOS} and {\SION}. Our results thereby establish the capabilities of O-$K$ RIXS in studying dispersive magnetic and electronic excitations in iridates and manifest the strong Ir $5d$-O $2p$ hybridization in these materials. In addition, dimensionality and temperature dependence of the excitations in {\SIOS} and {\SION} indicate an electron-hole exciton nature of the sharp dispersive mode [E in Fig. 1(g)].

The {\SIOS} and {\SION} single crystals used in this study were grown by the flux method \cite{cao214, cao327}. The x-ray absorption (XAS) and RIXS measurements were carried out at the ADRESS beamline of the Swiss Light Source at the Paul Scherrer Institut \cite{beamline}, with both $\pi$ and $\sigma$ polarizations [Fig. 1(d)]. The scattering angle was set to $2\theta_s = 130^{\text{o}}$, by which a substantial area of the first Brillouin zone is accessible [pink circle in Fig. 1(b)]. The measurements were performed along two high symmetry directions [H, 0] and [H, H] in tetragonal notation with $a=b\approx 3.9$ {\AA}. The energy resolution for the RIXS measurements was set to $65$ meV. The in-plane momentum $q_{\parallel}$ can be tuned continuously by rotating the sample and thereby changing the angle ($\delta$) [Fig. 1(d)].

Figures 1(e) and 1(f) show the XAS spectra for {\SIOS} and {\SION} measured at a grazing-incidence angle ($\delta = -45^{\text{o}}$) [$\mathbf{Q} = 0.43(\pi, 0)$] using $\pi$ and $\sigma$ polarizations. The polarization dependence is determined by the hybridization between Ir $5d$ $t_{2g}$ and O $2p$ orbitals and the matrix elements of the O $1s-2p$ dipole transitions.  The $\sigma$ polarization probes Ir $5d$ $yz/zx$ ($xy$) orbitals hybridized with $2p_x/2p_y$ orbitals of apical (plane) oxygens. The $\pi$ polarization predominantly favors the Ir $5d$ $yz/zx$ orbitals via plane-oxygen $2p_z$ orbitals \cite{bjkim_08,chang_15,marco_14}. After having determined the $K$-edge resonant energies for plane and apical oxygens [$yz/zx_\text{A}$, $xy_{\text{P}}$ and $yz/zx_\text{P}$ in Figs. 1(e) and 1(f)], we have performed incident energy dependent RIXS measurements covering these resonant energies. The results measured with $\pi$ polarization are shown in Figs. 1(g) and 1(h). Clear Raman peaks A/A', E, C/C' observed below $1$ eV are identified as single magnons, electron-hole exciton across the charge gap, and spin-orbit excitons, respectively [Fig. 1(c)]. B/B' and BM resolved via multi-gaussian fitting are spin-orbit excitons ``dressed'' by magnons, and bimagnons, respectively [Fig. 1(c)].

\begin{figure}[htbp]
\includegraphics[width=8.5cm]{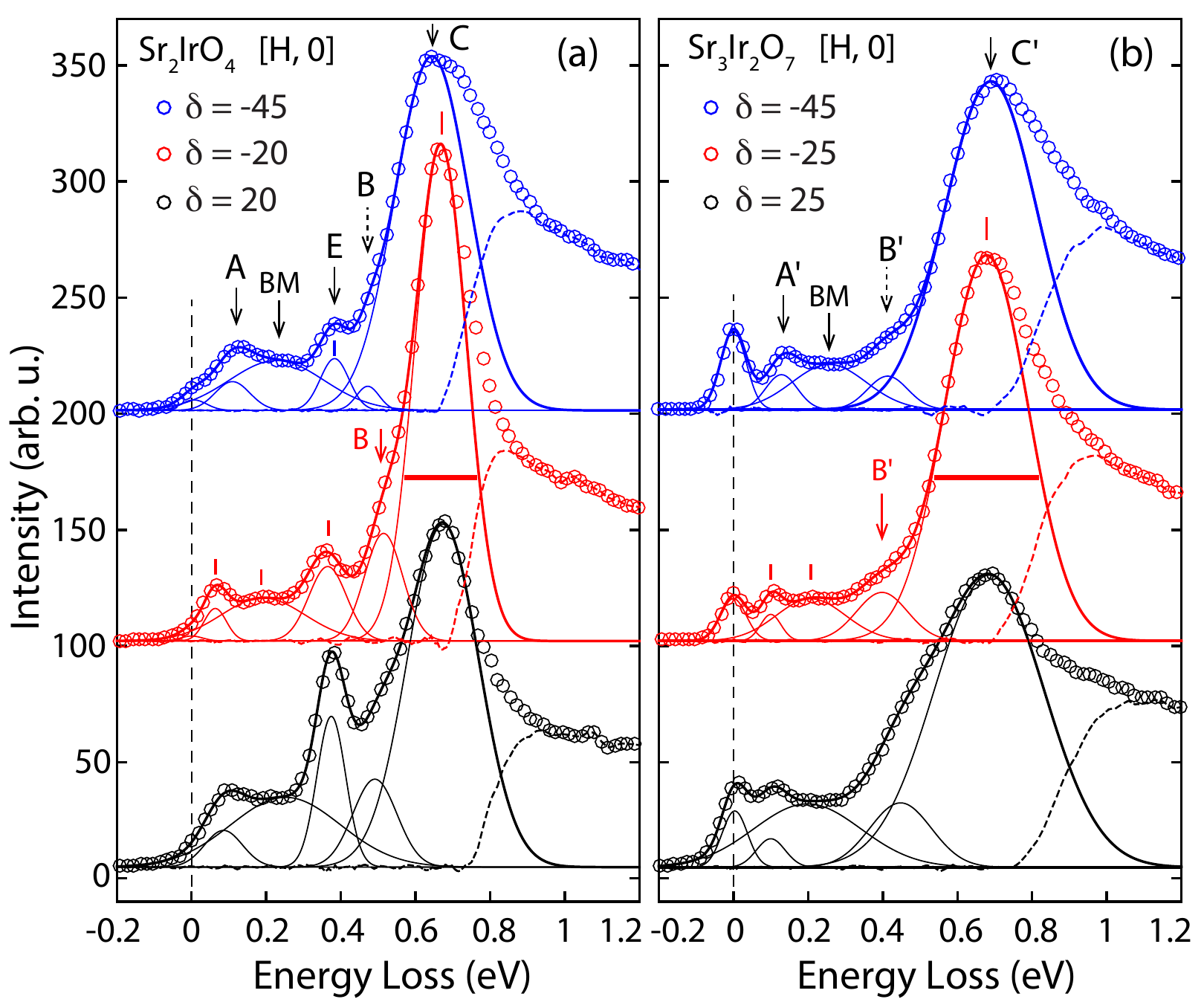}
\caption{(color online). Multi-gaussian fitting of the RIXS spectra for (a) {\SIOS} and (b) {\SION}. The spectra at three different $\delta$s are shown as colored open circles. From low to high energy loss, the Gaussian peaks (solid curves) denote elastic peak (centered at zero energy loss), single magnon (A/A'), bimagnon (BM), electron-hole exciton (E), excitonic quasiparticle (B/B') and spin-orbit exciton (C/C'). The dashed curves are the residual weight of the fitting.}
\label{fig.F2}
\end{figure}

Figure 2 shows the fitting of selected spectra from grazing incidence [$\delta = -45^{\text{o}}$, $0.43(\pi, 0)$] to normal incidence [$\delta = 25^{\text{o}}$, $0.25(\pi, 0)$] for {\SIOS} and {\SION} using multiple Gaussians. The systematic fitting resolve all the collective modes below $1$ eV and show their incident-angle dependent cross sections. To identify the nature of these collective modes, we have measured the momentum dependence of the RIXS spectra along [H, 0] and [H, H] for both {\SIOS} and {\SION} [Fig. 3], with the incident photon energy tuned to the resonant energy $yz/zx_{\text{P}}$ that enhances all the Raman modes. The spectra shown in Figs. 2 and 3 are normalized to the charge transfer excitations between $5$ and $11$ eV \cite{SI}. It is remarkable that A and A' show clear dispersions with energy gaps around $\mathbf{Q} = (0, 0)$. The extracted dispersions from the fitting coincide with the magnon dispersions measured by Ir-$L_3$ RIXS for both {\SIOS} and {\SION} [Fig. 4] \cite{jkim_214, bjkim_14ncom, jkim_327,marco_327}. Without any doubt, A and A' can be attributed to single magnons, proving that O-$K$ RIXS is sensitive to single magnons in the iridates with strong SOC in $5d$ shell and broken inversion symmetry at oxygen sites \cite{Brink_15}. The large magnon gap ($\sim 90$ meV) in {\SION} was explained to be driven by strong magnetic anisotropy associated with SOC \cite{jkim_327,marco_327}. The smaller magnon gap in {\SIOS} ($\sim 40$ meV) is consistent with an Ir-$L_3$ RIXS report \cite{bjkim_14ncom} and was described by including an $XY$ anisotropic term \cite{vale_15, vale_17} into the Heisenberg model described in Ref. \cite{jkim_214}. Besides the dispersions, we find that the single magnons persist in the whole momentum space measured with undiminished spectral weight [Figs. 2 and 3], therefore sorting out the momentum dependence of the magnon intensity, which was not settled in previous studies \cite{Brink_15,liu_15}.

\begin{figure}[htbp]
\includegraphics[width=8.5cm]{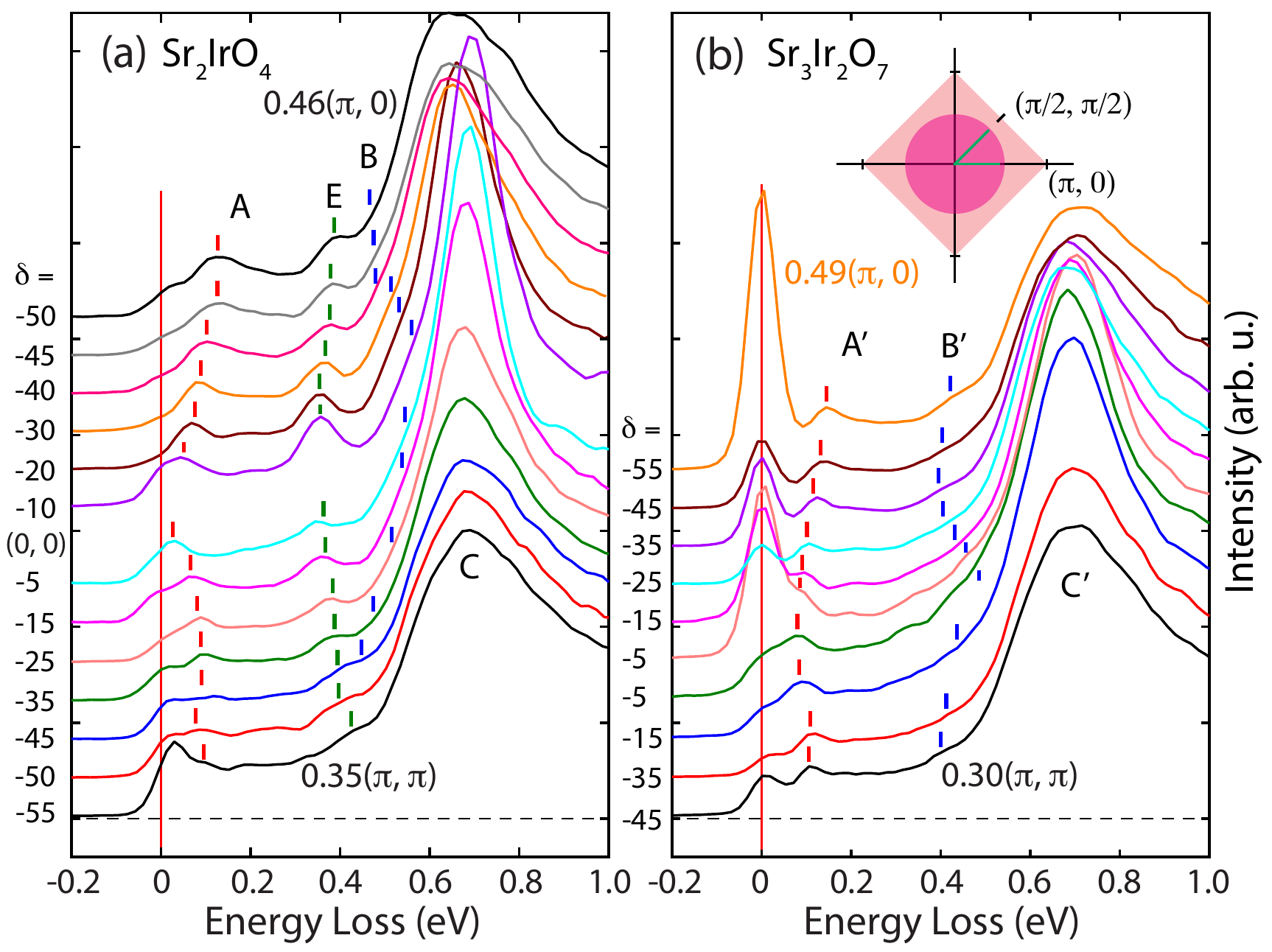}
\caption{(color online). Momentum-dependent RIXS spectra along [H, 0] and [H, H] directions for (a) {\SIOS} and (b) {\SION}. The inset in (b) illustrates the reciprocal space, where the pink diamond is the magnetic Brillouin zone and the pink filled circle the momentum coverage by O $K$ RIXS. The green lines are the [H, 0] and [H, H] high symmetry directions for RIXS measurements. The red vertical lines mark the zero energy loss positions. The red, green and blue markers indicate the fitted peak energies of the single magnons (A/A'), the electron-hole excitons (E) and the excitonic quasiparticles (B/B'), respectively.}
\label{fig.F3}
\end{figure}

The fitting in Fig. 2 reveals a broad spectral structure at $\sim 200$ meV for both {\SIOS} and {\SION} [blue squares in Fig. 4], which show similar dispersions to the corresponding single magnons. This feature can be attributed to bimagnons, because (1) their energy scales are consistent with the bimagnons  (160 meV for {\SIOS} and 185 meV for {\SION}) measured by Raman scattering \cite{gretarsson_raman} and (2) their dispersions are in agreement with those measured by Ir-$L_3$ RIXS \cite{bjkim_14ncom, jkim_327, marco_327}. Note that the bimagnons of cuprates measured by O-$K$ RIXS are non-dispersive \cite{valentina_OK_1, valentina_OK_2}. The presence of both single magnons and bimagnons indicates that O-$K$ RIXS is a powerful spectroscopic method for studying the magnetic dynamics of iridates.

\begin{figure}[htbp]
\includegraphics[width=8.5cm]{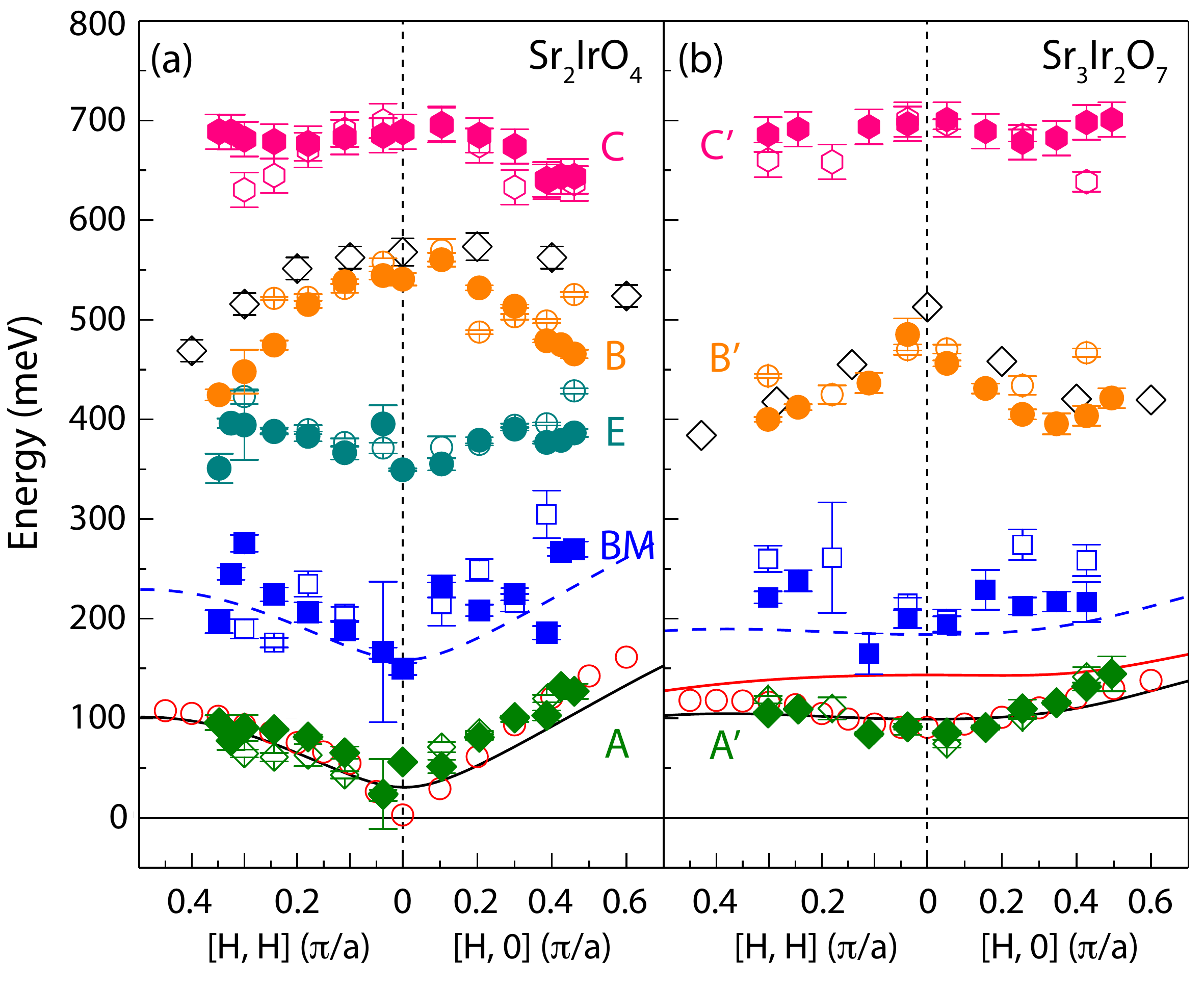}
\caption{(color online). Dispersions of the collective modes in (a) {\SIOS} and (b) {\SION} probed by O-$K$ RIXS. From low to high energy loss, the green diamonds, blue squares, cyan circles, orange circles and pink hexagons denote the single magnons (A/A'), bimagnons (BM), electron-hole excitons (E), spin-orbit excitons creating magnons (B/B') and spin-orbit excitons (C/C'), respectively. The solid and open symbols are extracted from the spectra with $\delta<0$ and $\delta>0$, respectively [Figs. 1 and 2]. The red open circles are the single magnon energies extracted from previous Ir-$L_3$ RIXS measurements \cite{jkim_214, marco_327}. The black open diamonds denote the dispersions of the excitonic quasiparticles extracted from Ref. \cite{bjkim_14ncom, stefano_thesis}. The black curve in (a) is a theoretical fit of the magnon dispersion using a Heisenberg model reported in \cite{vale_15, vale_17}. The black and red curves in (b) are transverse and longitudinal magnon dispersions for {\SION} using the quantum-dimer model reported in \cite{marco_327}. The blue dashed curves are the expected lower bound of the bimagnon continuum.}
\label{fig.F4}
\end{figure}

In addition to the magnetic excitations, O-$K$ RIXS also reveals the dispersive spin-orbit excitons (B and C) in {\SIOS} [Fig. 1(c)] as observed in previous Ir-$L_3$ RIXS experiments \cite{bjkim_14ncom, gretarsson_la214}. C can be easily attributed to a spin-orbit exciton mode (at $\sim 0.7$ eV) that exhibits minor dispersion and strong intensity \cite{bjkim_14ncom, gretarsson_la214}. The mode B finds direct correspondence with the excitonic quasiparticle (spin-orbit exciton ``dressed'' by magnons) reported in Ref. \cite{bjkim_14ncom}, because of their coincidence in dispersion [Fig. 4(a)] and similarities in band width [Figs. 2(a) and 4(a)]. Furthermore, the intensity of B increases while approaching normal incidence from grazing incidence [Fig. 2(a)]. This incident-geometry dependence of intensity is also consistent with that for the excitonic quasiparticles \cite{bjkim_14ncom} and therefore corroborates this identification. The spin-orbit exciton modes B and C arise from the electronic transitions from {\jeffth} quartet to {\jeffh} doublet, with B arising mainly from the $|j = 3/2, j_z = \pm 3/2>$ doublet, while C bears mostly $|j = 3/2, j_z = \pm 1/2>$ character. The energy difference between B and C ($\Delta_{\text{BC}}\approx 150$ meV) reflects the splitting of the \jeffth band, which is caused by the tetragonal distortion of the IrO$_6$ octahedra \cite{bjkim_14ncom}.

A similar analysis can be applied to the collective modes B' and C' in {\SION}. The mode B' is the excitonic quasiparticle in {\SION} and C' the spin-orbit exciton mode at higher energy ($\sim 0.7$ eV). B' and C' show similar dispersions to B and C, indicating similar origin and behaviors. However, the energy widths of B' and C' are larger than that for B and C, as evidenced from our fitting analysis in Fig 2. This larger bandwidth is caused by the increase in dimensionality (more adjacent IrO$_6$ octahedra) in {\SION}  \cite{dimensionality_mit}. The energy of B' is obviously lower than B, meaning that the splitting of the {\jeffth} states in {\SION} is larger ($\Delta_{\text{B'C'}}\approx 200$ meV), indicating a larger tetragonal distortion field in {\SION} \cite{gretarsson_213, bjkim_14ncom}. 

The sharp collective mode E below $0.4$ eV is an intriguing feature whose origin is still under debate \cite{bjkim_14ncom, jt_prl}. It was firstly suggested to be an electron-hole exciton at the edge of the electron-hole continuum, which has a threshold at about $0.41$ eV \cite{bjkim_14ncom, gretarsson_213, moon_09} [Fig. 1(c)]. Alternatively, it was explained as driven by a Jahn-Teller effect with strong SOC \cite{jt_prl}. In our measurements, E decreases in intensity with increasing temperature and disappears at $300$ K [SFig. 7 of the supplemental material] \cite{SI}. This is consistent with the decrease and broadening of the electron-hole continuum threshold at high temperatures as indicated by optical conductivity measurements \cite{moon_09}. The absence of E in {\SION} [Figs. 2 and 3] can also be unified in this exciton picture, as the much smaller charge insulating gap ($\sim 130$ meV) and the broader bandwidth of {\SION} could smear this exciton and push it to lower energy ($\sim 100$ meV) \cite{327_stm, dimensionality_mit}.

The excitonic quasiparticles B and B' ($|j = 3/2, j_z = \pm 3/2>$ to $|j = 1/2, j_z = \pm 1/2>$) deserve special attention. The propagation of the excitonic quasiparticles in the antiferromagnetic context creates flipped spins along their hopping path \cite{bjkim_14ncom}. Moreover, B and B' are magnetic $dd$ excitations in nature. From the selection rules for dipolar transitions, $\Delta j_z = 0, \pm1$, it follows that B/B' contains only spin-1 magnetic components, which usually rotates the photon polarization by $\pi/2$, the same as the single spin-flip excitations. This is consistent with previous theoretical calculations \cite{Brink_15}, which also generate magnetic $dd$ excitations within $5d$ $t_{2g}$ orbitals of {\SIOS}. Therefore, we conclude that the excitonic quasiparticles observed by O-$K$ RIXS are magnetic orbital excitations ``dressed'' by magnons. The spin-orbit exciton mode C/C' consists of both non-magnetic and magnetic components.

The capability of O-$K$ RIXS in probing magnetic excitations in iridates is driven by the strong SOC in the Ir $5d$ shell and hybridization between Ir $5d$ $t_{2g}$ and O $2p$ orbitals because spin is not a good quantum number in Ir $5d$ $t_{2g}$ states, indirect electronic transitions between Ir $5d$ $t_{2g}$ orbitals and O $1s$ states via strong O $2p$-Ir $5d$ hybridization allow flipping of the {\jeffh} pseudospin and changing of the {\jeffth} states, in which the former induce single magnons and the latter magnetic (and non-magnetic) $dd$ excitations. The role of the space inversion symmetry breaking at oxygen sites need further O-$K$ RIXS studies on iridate systems with different Ir-O-Ir bond angles.

In conclusion, our results on {\SIOS} and {\SION} and their comparison with previous Ir-$L_3$ RIXS results have established the capability of O-$K$ RIXS in probing single magnons and excitonic quasiparticles over a substantial part of the first Brillouin zone of layered iridates, which host strong SOC-entangled states in $5d$ orbitals. O-$K$ RIXS has high energy resolution for studying most 5$d$ TM oxides \cite{note}. Since O-$K$ RIXS usually generates strong RIXS response within a small attenuation length, its unique capabilities in probing elementary excitations provide a new method for investigating 5$d$ TM oxides, especially for artificial thin films, heterostructures and superlattices.

\begin{acknowledgments}
The work at PSI and EPFL is supported by the Swiss National Science Foundation through its Sinergia network Mott Physics Beyond the Heisenberg Model (MPBH). The work in London is supported by the UK Engineering and Physical Sciences Research Council (Grants No. EP/N027671/1 and No. EP/N034694/1). X. L., V. B. and P. O.-V. acknowledge financial support from the European Community's Seventh Framework Program (FP7/2007-2013) under grant agreement NO. 290605 (COFUND: PSI-FELLOW). J.P. and T.S. acknowledge financial support through the Dysenos AG by Kabelwerke Brugg AG Holding, Fachhochschule Nordwestschweiz, and the Paul Scherrer Institut. J. P. acknowledge financial support by the Swiss National Science Foundation Early Postdoc. Mobility fellowship project number P2FRP2$\_$171824. The work at PSI is partially funded by the Swiss Nationtional Science Foundation through the D-A-CH program (SNSF Research Grants No. 200021L 141325). H. M. R. acknowledges financial support from SNSF grant 200020-166298.
\end{acknowledgments}

\renewcommand{\figurename}{SFigure}
\setcounter{figure}{0}

\newpage
\section*{SUPPLEMENTARY MATERIAL} 

\subsection{Resonant Inelastic X-ray Scattering at Transition-Metal $L$ Edge and Oxygen $K$ edge}

Resonant inelastic x-ray scattering (RIXS) is a photon-in-photon-out scattering technique suitable for measuring elementary excitations of condensed matter systems \cite{rmp_rixs}. In RIXS, the incident photon energy is tuned to an x-ray absorption edge of certain element to enhance the cross sections for specific elementary excitations \cite{rmp_rixs}. At the transition-metal (TM) $L_{2,3}$ edges, the incident photons excite a 2$p$ core electron (with spin $\sigma$) into the outer $d$ shell, leaving a core hole in the 2$p$ state. Because of the strong spin-orbit coupling (SOC) in TM 2$p$ shell, SOC-entangled $2p_{\frac{3}{2}, \frac{1}{2}}$ states are formed, where spin is no longer a good quantum number \cite{ament_09}. In the subsequent relaxation process, an electron with spin $-\sigma$ can therefore fill the core hole. The net result of this RIXS process is a spin-flip excitation. Therefore, single spin-flip excitations are directly allowed at TM-$L$ edge, driven by the strong SOC in the 2$p$ core level \cite{ament_09, LB_09, LB_10}. TM-$L$ RIXS is usually performed at the $L_3$ edge ($2p_{\frac{3}{2}}$) \cite{tacon_11, nature_12, kejin_13, jkim_214, bjkim_14ncom, jkim_327, marco_327}.

Different from TM-$L_3$ edge, oxygen-$K$ (O-$K$) edge involves O $1s-2p$ electronic transitions. Since SOC is absent at $s$ level, single spin-flip excitations are assumed to be forbidden at O-$K$ edge \cite{rmp_rixs}, as supported experimentally by previous RIXS reports on various 3$d$ and 4$d$ TM oxides \cite{marco_14, chang_15, valentina_OK_1, valentina_OK_2}. However, considering the strong SOC in 5$d$ levels of various 5$d$ TM oxides such as iridates and strong hybridyzation between TM 5$d$ orbitals and O 2$p$ orbitals \cite{jackeli_prl, review15, bjkim_08}, exploring single spin-flip excitation via O-$K$ RIXS could be possible. A theoretical proposal from Kim {\it et al.} \cite{Brink_15} suggested that single magnons and magnetic orbital excitations could be allowed in {\SIOS}, because of the strong SOC in Ir 5$d$ $t_{2g}$ orbitals and the broken space inversion symmetry at oxygen sites. Despite this, a conclusive experimental study demonstrating the capability of O-$K$ RIXS in probing single spin-flip excitations is still lacking.

Towards this aim, we have performed O-$K$ RIXS measurements on the archetypical iridates {\SIOS} and {\SION} \cite{jkim_214, jkim_327, jackeli_prl, review15, bjkim_08}. We have shown our key discoveries in the main text. Our results have unambiguously shown that O-$K$ RIXS is sensitive to single magnon and excitonic quasiparticles in the spin-1 channel. In this supplemental material, we show more detailed data to complement our studies on {\SIOS} and {\SION} using O-$K$ RIXS.

~\\

\subsection{Incident Photon Energy Dependent RIXS at Grazing Incidence}

\begin{figure}[htbp]
\includegraphics[width=8.5cm]{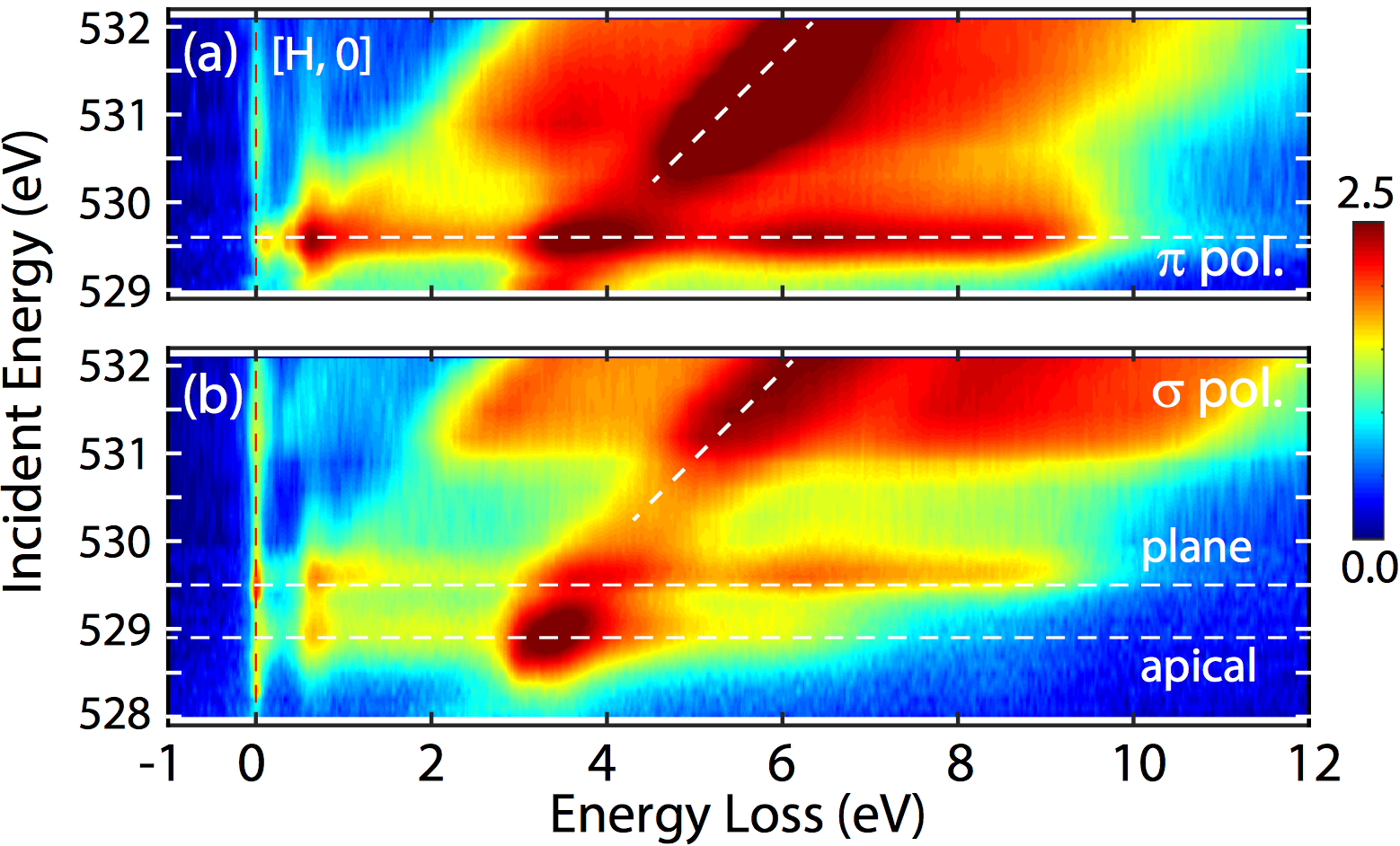}
\caption{Incident photon energy dependent elementary excitations of {\SIOS}, measured near the O-$K$ edge with $\pi$ and $\sigma$ polarizations. The measurements were performed with $\delta=-45^{\text{o}}$ along tetragonal [H, 0] direction. The diagonal white dashed lines mark the fluorescence lines. The resonant energies corresponding to the absorption peaks are marked by horizontal white dashed lines. The two white dashed lines in (b), $\sim528.8$ eV and $\sim529.6$ eV, denote the resonant peaks for apical and plane oxygen, respectively.}
\label{fig.F1}
\end{figure}

In Fig. 1 of the letter, we have shown incident photon energy ($E_i$) dependent RIXS in the low energy loss region ([-0.5, 2] eV) for $\pi$ polarization at a grazing incident angle ($\delta = -45^{\text{o}}$). Here we show in SFigs. 1 to 4 the $E_i$-dependent RIXS maps and selected spectra measured at resonant energies for both $\pi$ and $\sigma$ polarizations in a wide energy loss range ([-1, 12] eV). The $E_i$ ranges cover the O-$K$ absorption peaks [Figs. 1(c) and 1(d) of main text] associated with the Ir $t_{2g}$ $yz/zx$ and $xy$ orbitals \cite{marco_14, chang_15}. The white dashed lines mark the resonant absorption energies.

SFig. 1 depicts the whole $E_i$-dependent RIXS map for {\SIOS}.  Besides the Raman features marked in Fig. 1(e), excitations at $\sim3.5$ eV, $6.5$ eV and $8.8$ eV [SFig. 1(a)] containing Raman components have also been observed near the resonant energy associated with the Ir $t_{2g}$ $yz/zx$ orbitals ($\sim529.6$ eV) using $\pi$ polarized photons. The excitations at $\sim3.5$ eV can be attributed to the main O $2p$ decay channel with strong signature of the $dd$ excitations between Ir $t_{2g}$ and $e_g$ orbitals observed at O-$K$ edge due to the strong O$2p-$Ir$5d$ hybridization. The excitations above $6$ eV can be ascribed to charge transfer excitations. Above the resonant energy, strong fluorescence appears at energy loss above $2$ eV.

SFig. 1(b) shows the RIXS energy map measured with $\sigma$ polarization between $E_i=528$ and $532$ eV. Corresponding to the two resonant absorption peaks in Fig. 1(c), the RIXS map in SFig. 1(b) also shows two resonant energies enhancing different excitations. Both energies enhance the spin-orbit exciton at $\sim 0.7$ eV but fail to reveal the clear Raman features A ($\sim 0.1$ eV) and E ($\sim 0.4$ eV), which have been identified as the single magnon and an exciton across the charge gap as explained in the main text. 

\begin{figure}[htbp]
\includegraphics[width=8.5cm]{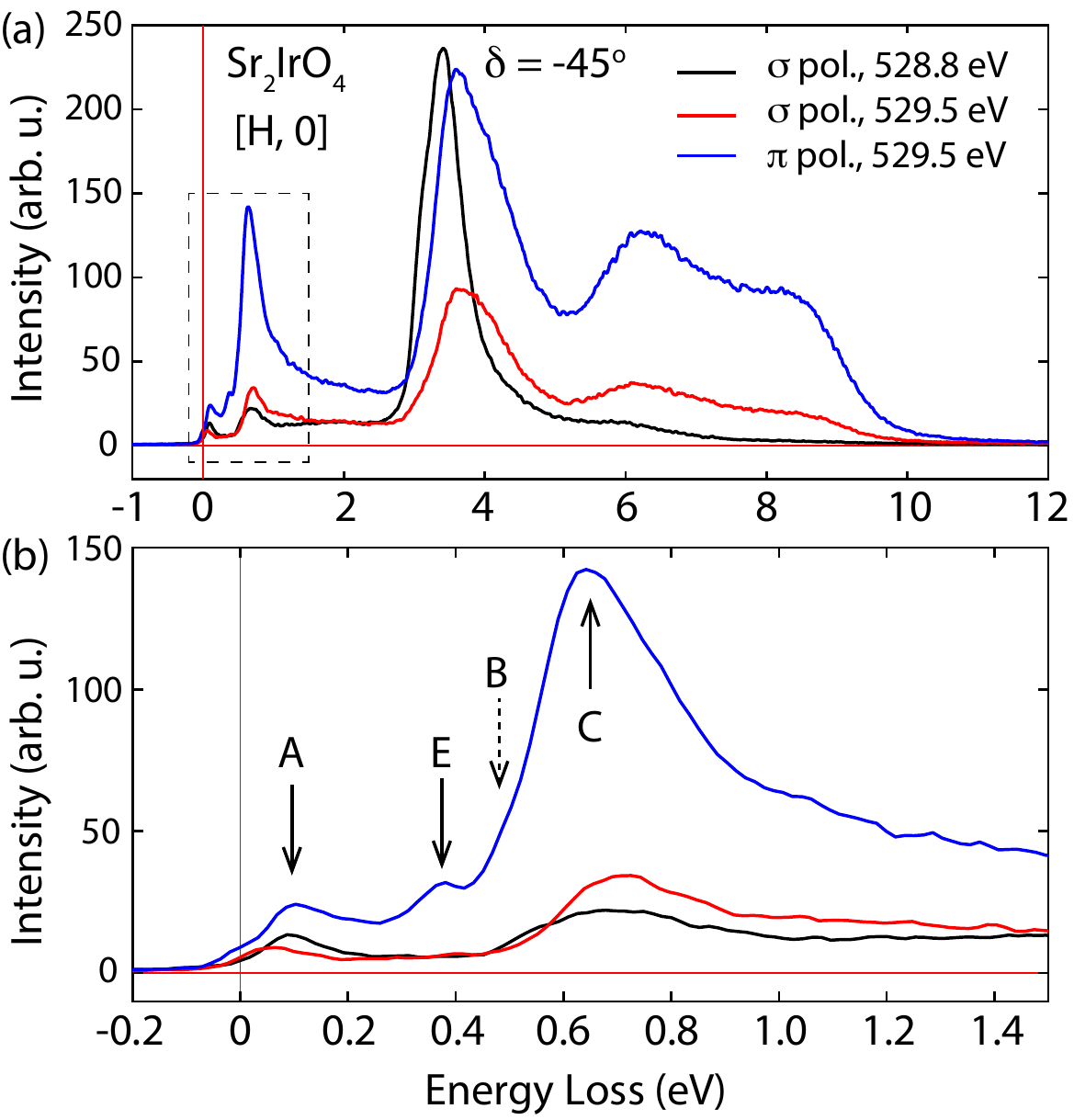}
\caption{Polarization-dependent RIXS spectra measured at resonant absorption peaks for {\SIOS}. The corresponding incident energies are marked by horizontal white dashed lines in SFig. 1 and green and red ticks in Fig. 1(c). (a) shows the full energy loss region, while (b) is a zoomed-in view of the rectangular region in (a). Raman features A, E, B, and C are marked by arrows.}
\label{fig.F2}
\end{figure}

To show polarization dependence quantitatively, we compare the RIXS spectra collected at the three resonant energies [white dashed lines in SFig. 1] in SFig. 2. It is clear that all the Raman features are greatly enhanced at the plane-oxygen resonant energy with $\pi$ polarization [$yz/zx_{\text{P}}$ in Fig. 1(c)], which probes Ir $d_{yz}/d_{zx}$ orbitals through their hybridization with $p_z$ orbitals of plane-oxygen ions \cite{marco_14}. This could be understood considering that all these collective modes propagate within the IrO$_2$ plane \cite{jkim_214, bjkim_14ncom}. The charge transfer excitations are also enhanced at $yz/zx_{\text{P}}$ ($E_i \approx 529.6$ eV, $\pi$ polarization), and they can also be clearly observed when exciting at the plane-oxygen energy with $\sigma$ polarization that probe the Ir $t_{2g}$ $xy$ orbital [$xy_{\text{P}}$ in Fig. 1(c)]. In contrast, the charge transfer excitations are much weaker in the spectrum measured at the apical-oxygen resonant energy $yz/zx_{\text{A}}$ ($E_i \approx 528.8$ eV, $\sigma$ polarization). Therefore, we speculate that the charge transfer process occurs mostly within the IrO$_2$ plane, in agreement with the single-layer structure of {\SIOS}.

\begin{figure}[htbp]
\includegraphics[width=8.5cm]{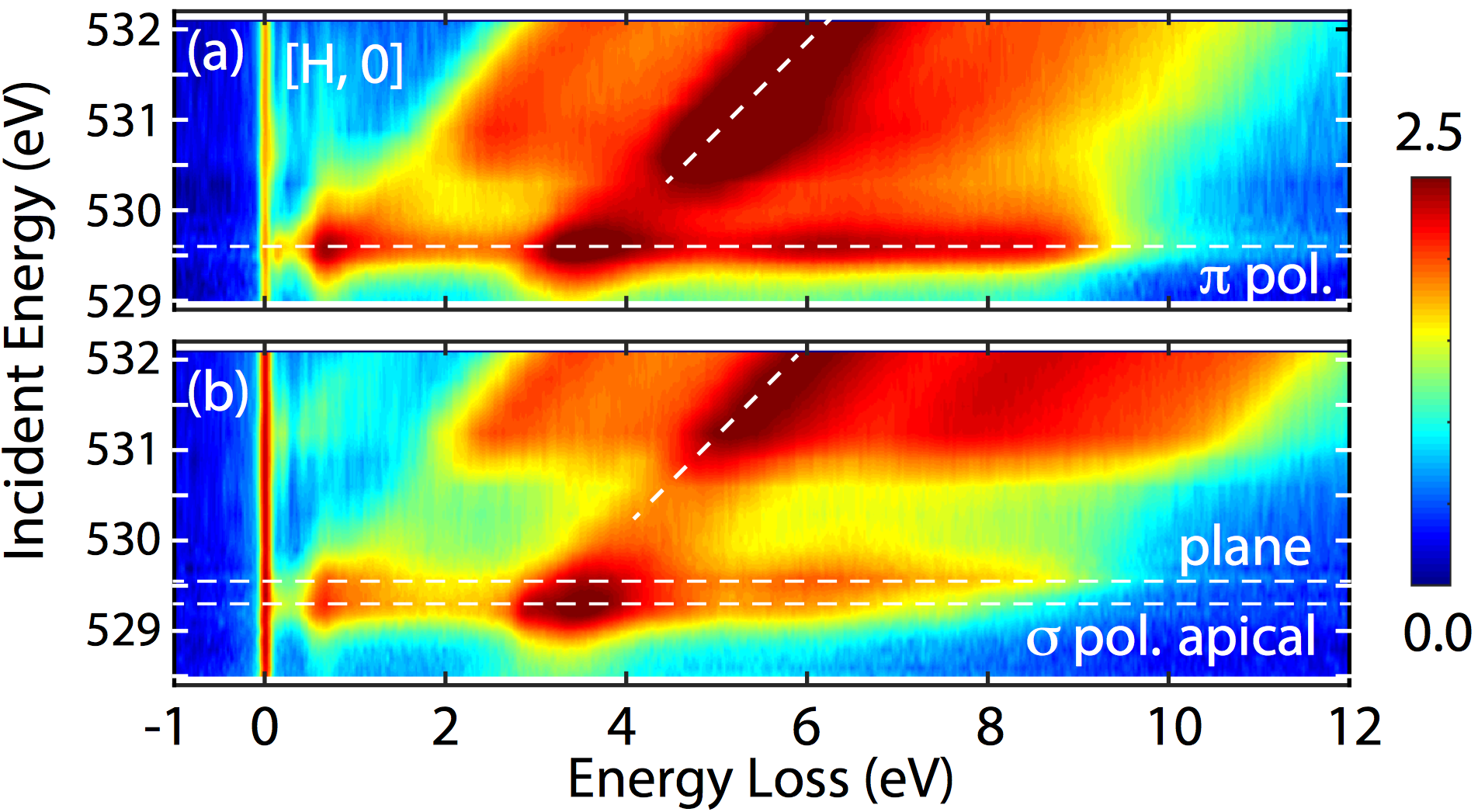}
\caption{$E_i$-dependent elementary excitations for {\SION}, measured near the O-$K$ edge with $\pi$ and $\sigma$ polarizations along [H, 0] and [H, H] directions, with $\delta = -45^{\text{o}}$. The diagonal white dashed lines mark the corresponding fluorescence lines. The resonant energies are marked by horizontal white dashed lines. The two white dashed lines at $\sim 529.3$ eV and $\sim 529.5$ eV in (b) denote the resonant peaks for apical and plane oxygen, as shown in Fig. 1(d), respectively.}
\label{fig.F3}
\end{figure}

The same data analysis and interpretations can be applied to the polarization dependence of the grazing-incident spectra for {\SION}. All the collective excitations favor the plane-oxygen resonant energy with $\pi$ polarization [$yz/zx_{\text{P}}$ in Fig. 1(d)] [SFigs. 3 and 4]. Because of the bilayer structure, the difference between the resonant energies $yz/zx_{\text{P}}$ and $yz/zx_{\text{A}}$ decreases to a much smaller value ($\sim 0.2$ eV) than that in {\SIOS}, as shown in Fig. 1(d) and SFig. 3. The bilayer structure also increases the charge transfer excitations (above 5 eV) probed via the apical-oxygen resonant energy [$yz/zx_{\text{A}}$ in Fig. 1(d)] [SFig. 4(a)].

\begin{figure}[htbp]
\includegraphics[width=8.5cm]{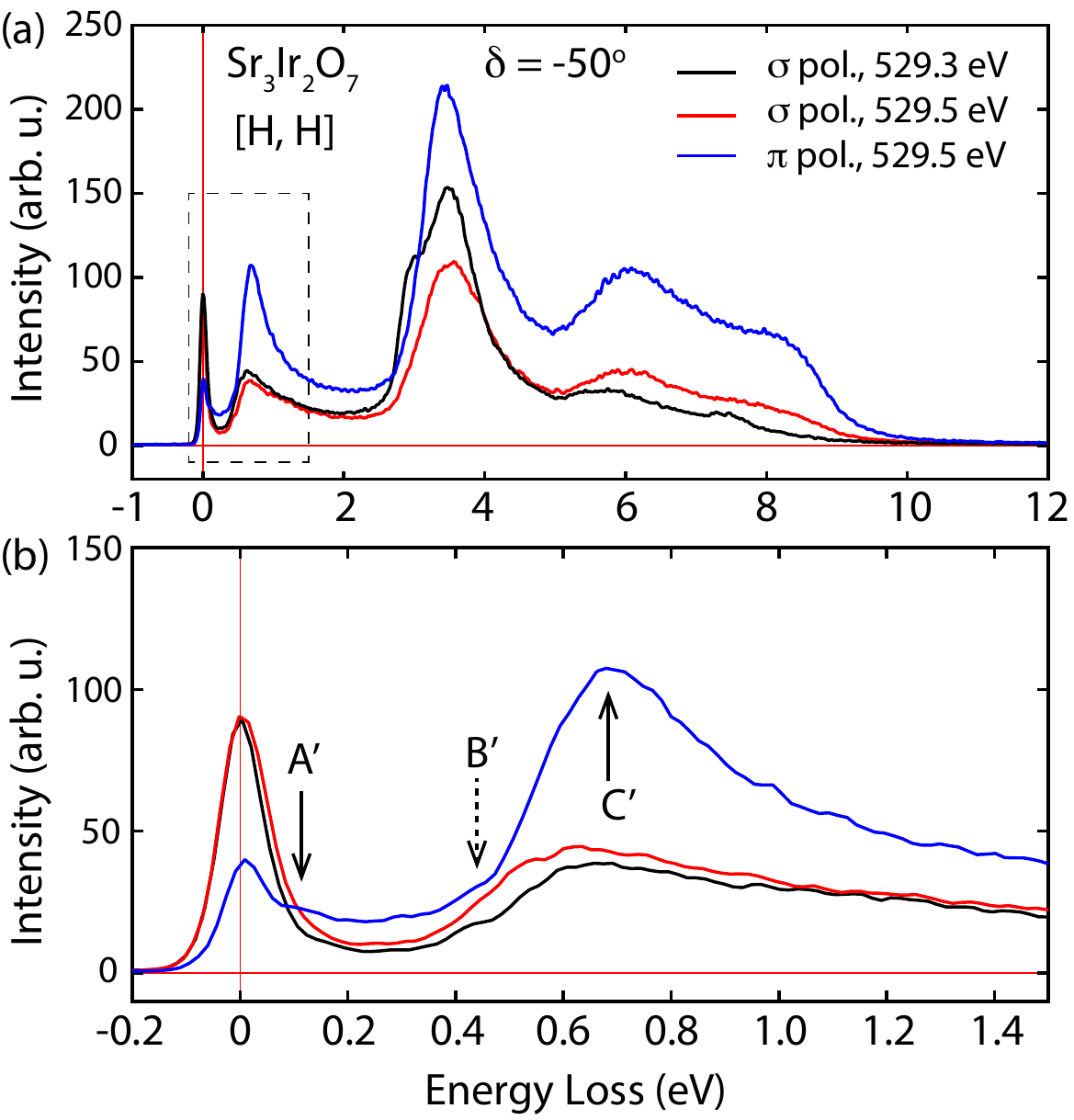}
\caption{(a) Polarization-dependent RIXS at resonant absorption peaks for {\SION}. The corresponding incident energies are marked by horizontal white dashed lines in SFig. 3 and green and red ticks in Fig. 1d. (a) is for full energy loss region, while (b) is a zoomed-in view of the rectangular region in (a). The Raman features A', B', and C' are marked by arrows.}
\label{fig.F4}
\end{figure}

~\\

\subsection{Angular ($\delta$) Dependence of the RIXS spectra: Momentum Dispersion}

\begin{table}[htp]
\caption{Sample angle $\delta$ and its corresponding in-plane momentum, which is calculated with the scattering angle $2\theta_s = 130^{\text{o}}$, incident energy $E_i = 529.6$ eV and tetragonal lattice parameters $a = b = 3.89 {\AA}$ \cite{jkim_214, jkim_327}.}

\begin{tabular}{|c|c|c|c|}\hline
|$\delta$| ($^{\text{o}}$)      &$q_{//} (1/{\AA})$     \centering &$[H, 0] (\pi/a)$     \centering   &$[H, H] (\pi/a)$ \\
\hline
$0$       &$0$            &$0$        &$0$ \\
\hline
$5$       &$0.0423$   &$0.0524$     &$0.0370$  \\
\hline
$10$     &$0.0843$   &$0.1043$    &$0.0738$ \\
\hline
$15$     &$0.1256$   &$0.1555$     &$0.1100$   \\
\hline
$20$     &$0.1659$   &$0.2055$     &$0.1453$  \\
\hline
$25$     &$0.2051$   &$0.2539$     &$0.1795$  \\
\hline
$30$     &$0.2426$   &$0.3004$    &$0.2124$ \\
\hline
$35$     &$0.2783$   &$0.3446$     &$0.2437$   \\
\hline
$40$     &$0.3119$   &$0.3862$     &$0.2731$  \\
\hline
$45$     &$0.3431$   &$0.4248$    &$0.3004$ \\
\hline
$50$     &$0.3717$   &$0.4602$     &$0.3254$   \\
\hline
$55$     &$0.3975$   &$0.4921$     &$0.3480$  \\
\hline
$60$     &$0.4202$   &$0.5203$    &$0.3679$ \\
\hline
\end{tabular}
\label{table1}
\end{table}

\begin{figure}[htbp]
\includegraphics[width=8.5cm]{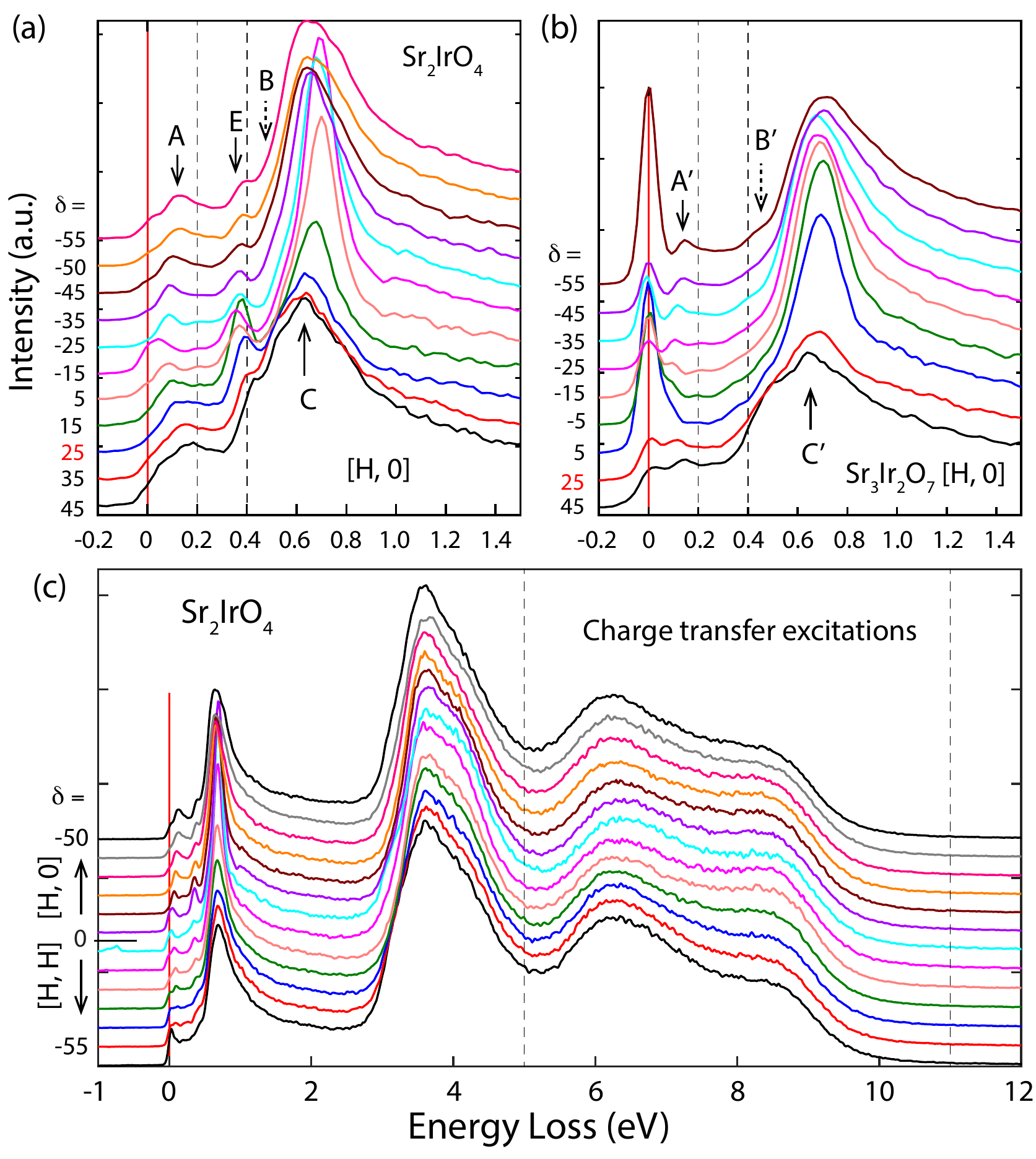}
\caption{Momentum dependence of the RIXS spectra for {\SIOS} and {\SION}. (a) Momentum-dependent RIXS along [H, 0] direction for (a) {\SIOS} and (b) {\SION}. The measurements were performed with both positive and negative $\delta$. (c) Full-energy RIXS spectra of {\SIOS} along [H, 0] and [H, H] directions. All the spectra in (a)-(c) are normalized to the charge transfer excitations between 5 and 11 eV, which is indicated by vertical dashed lines in (c).}
\label{fig.F5}
\end{figure}

The in-plane momentum can be tuned by changing the sample angles $\delta$ and $\phi$, where $\delta$ controls the magnitude of the projected in-plane momentum ($q_{\parallel}$) and the azimuthal angle $\phi$ determines the scattering plane. In our measurements, the momentum directions [H, 0] and [H, H] were reached by fixing $\phi$ at specific angles. To facilitate our measurements, we tune $\delta$ with a step of $5^{\text{o}}$. The corresponding in-plane momenta along [H, 0] and [H, H] are calculated and shown in Table I, in the unit of $(\pi/a)$.

SFigure 5 shows $\delta$ dependence of RIXS spectra. SFigs. 5(a) and 5(b) show RIXS spectra for {\SIOS} and {\SION} at positive and negative sides of $\delta$, which generate consistent dispersions for the collective modes [Fig. 4]. On the other hand, it is well known that RIXS cross section is sensitive to incident geometry. As shown in SFigs. 5(a) and 5(b), all the collective excitations are clearly observed in negative $\delta$ (close to grazing incidence), while they are difficult to be identified and distinguished from each other in grazing exit side ($\delta > 25^{\text{o}}$), where the single magnon (A and A') and the spin-orbit exciton at $\sim 0.7$ eV (C and C') become broader and less pronounced. However, the excitonic quasiparticles (B and B') are enhanced close to normal incidence ($\delta = 25^{\text{o}}$), which is consistent with previous results measured by Ir-$L_3$ RIXS \cite{bjkim_14ncom}. All the spectra in Fig. 2, 3 and SFig. 5 are normalized to the charge transfer excitations in the energy range of $[5, 11]$ eV, as indicated by vertical dashed lines in SFig. 5(c). 

\subsection{Comparison between O-$K$ RIXS and Ir-$L_3$ RIXS}

\begin{figure}[htbp]
\includegraphics[width=8.5cm]{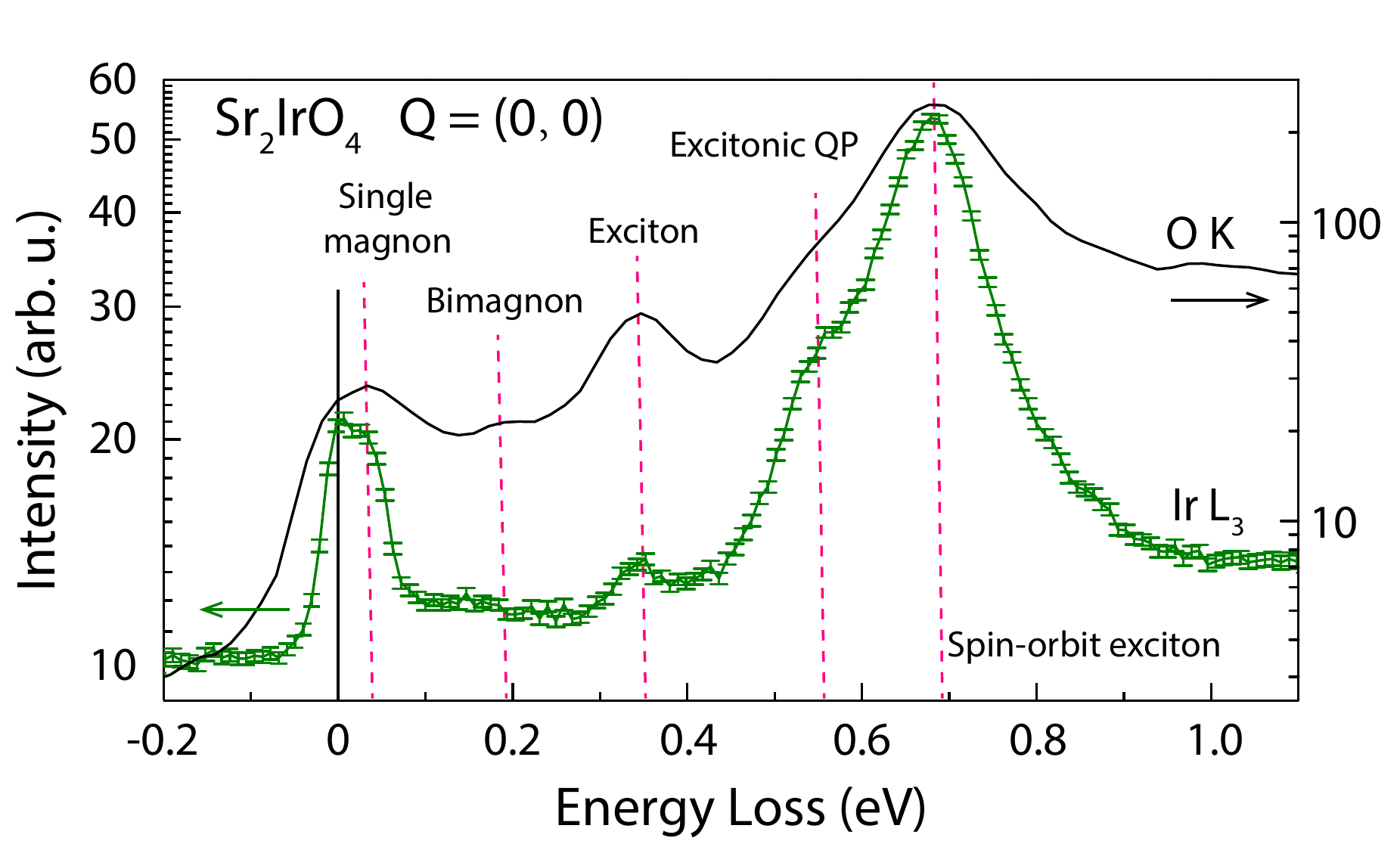}
\caption{Comparison of the Ir-$L_3$ and O-$K$ RIXS spectra measured at $Q = (0, 0)$ of {\SIOS}. The $y$ axis (intensity) is displayed in log scale. The vertical pink dashed lines mark the Raman features revealed at both edges. The spectrum measured by O-$K$ RIXS was collected in 80 minutes of counting time at SAXES, ADRESS beamline ($\Delta E \approx 65$ meV) with the flux at sample about $6\times 10^{13}$/s. The one for Ir-$L_3$ RIXS was measured in 150 minutes of counting time at 27-ID-B of Advanced Photon Source ($\Delta E \approx 35$ meV), with the flux at sample about $4\times 10^{13}$/s, roughly the same as that for O-$K$ RIXS.}
\label{fig.F6}
\end{figure}

A direct comparison between O-$K$ RIXS and Ir-$L_3$ RIXS in measuring the same momentum can provide a compelling evidence for the capability of O-$K$ RIXS in probing elementary excitations. For this purpose, we have measured the spectrum for $\textbf{Q} = (0, 0)$ of {\SIOS} using both O-$K$ and Ir-$L_3$ RIXS [SFig. 6]. To make the comparison clear, the spectral intensity in SFig. 6 is plotted in log scale and a constant background ($10$) is added to the Ir-$L_3$ spectrum. It is clear that both spectra host similar collective modes at same energies, as marked by pink dashed lines in SFig. 6.

The single magnon exhibits substantial intensity and forms a peak at $\sim 40$ meV overlapping with the elastic peak, consistent with recent Ir-$L_3$ results and the importance of the $XY$ magnetic anisotropy in {\SIOS} \cite{bjkim_14ncom, vale_15}. The feature at about $200$ meV could be attributed to bimagnon \cite{gretarsson_raman}, which is much stronger at O-$K$ edge than that in Ir-$L_3$ edge [Fig. 3]. We note that the bimagnons are as strong as the single magnons at the O-$K$ edge. This may cause some difficulties in extracting the single magnons, which can be alleviated by better energy resolution, as has recently become available \cite{weisheng_13}.The spin-orbit exciton at $\sim 0.7$ eV and the excitonic quasiparticle at $\sim 0.55$ eV were reported and interpreted in previous reports \cite{jkim_214, bjkim_14ncom}. The mode at $\sim 0.35$ eV is an exciton mode at the verge of the electron-hole excitation continuum across the insulating gap \cite{bjkim_14ncom, gretarsson_la214}. The coincidence of the collective mode energies between O-$K$ and Ir-$L_3$ RIXS spectra demonstrates that O-$K$ RIXS is capable of  measuring single magnons and excitonic quasiparticles.

\subsection{Temperature Dependence of the Elementary Excitations in {\SIOS} and {\SION}}

\begin{figure*}[htbp]
\includegraphics[width=12cm]{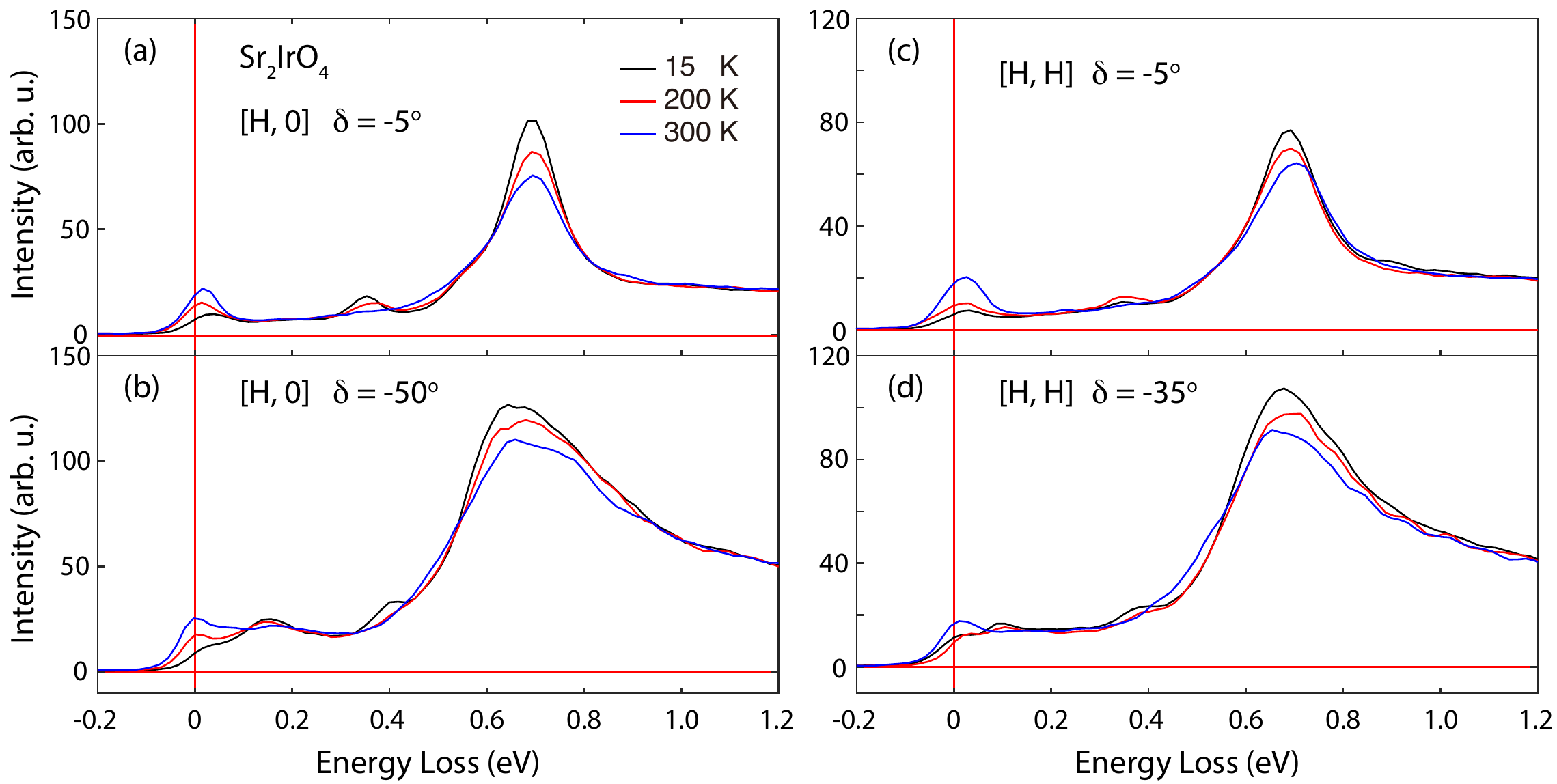}
\caption{Temperature dependence of magnetic excitations and spin-orbit excitons for {\SION} measured with $\delta = -45^{\circ}$ along (a) [H, 0] and (b) [H, H] directions. The spectra were collected at $T = 15, 100, 240$ and $300$ K. }
\label{fig.F8}
\end{figure*}

\begin{figure}[htbp]
\includegraphics[width=8.5cm]{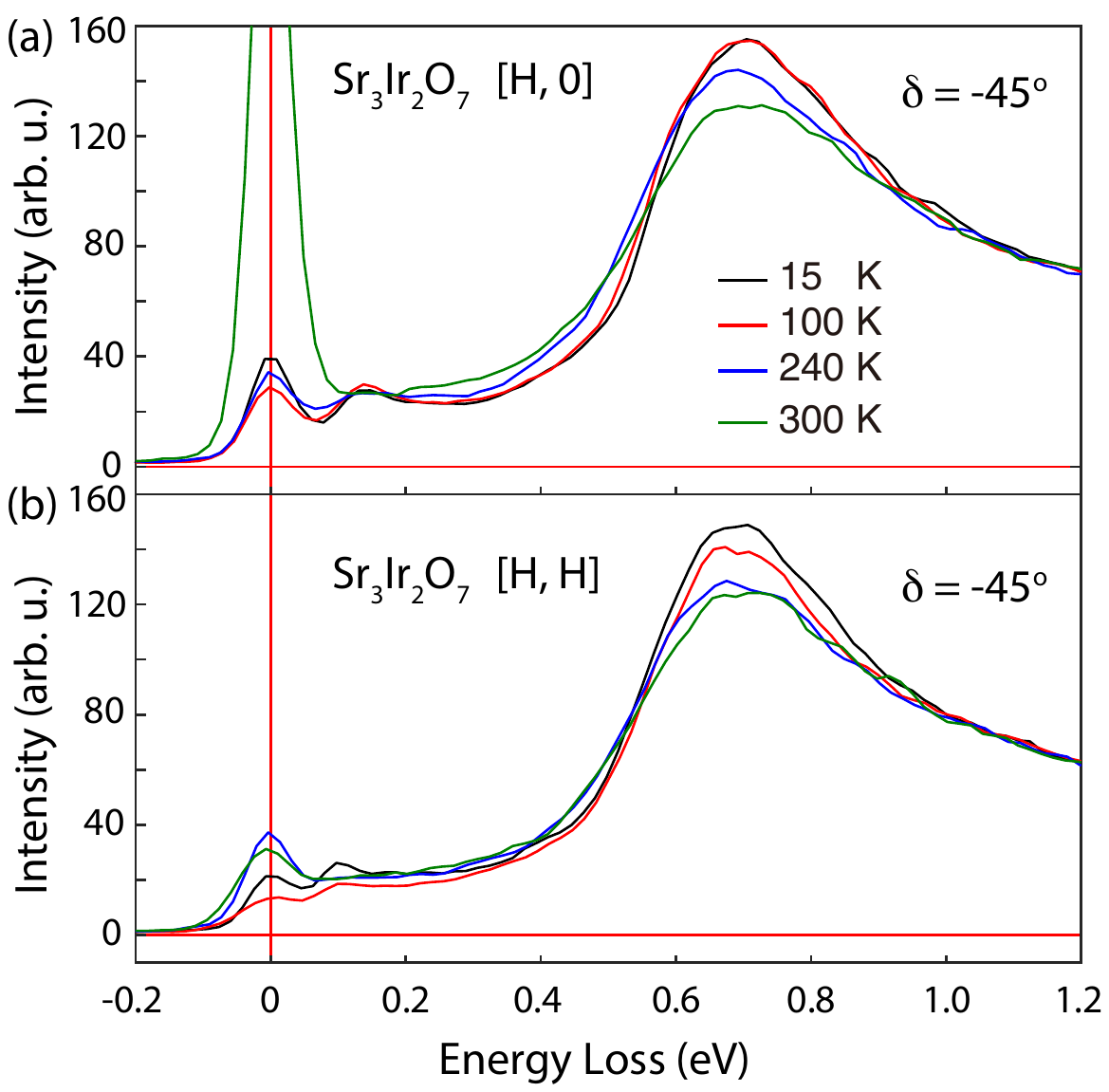}
\caption{Temperature dependence of magnetic excitations and spin-orbit excitons for {\SION} measured with $\delta = -45^o$ along (a) [H, 0] and (b) [H, H] directions. The spectra were collected at $T = 15, 100, 240$ and $300$ K. }
\label{fig.F9}
\end{figure}

SFigs. 7 and 8 depict the temperature dependence of the  elementary excitations below 1 eV for {\SIOS} and {\SION}. Upon increasing temperature for {\SIOS}, the elastic and quasielastic scattering are enhanced. The single magnons decrease in intensity and become less pronounced [SFig. 7]. The excitonic QP at $E \lessapprox 0.4$ eV [B in SFig. 2(b)] are heavily damped and disappear at 300 K, while the other excitonic QP at $\sim 0.5$ eV changes less and the main spin-orbit exciton at $\sim 0.7$ eV decrease in intensity by $\sim 20\% - 40\%$ [SFig. 8]. This indicates that the mode is either sensitive to itinerant electrons \cite{gretarsson_la214} or based on the in-plane antiferromagnetic structure of {\SIOS} ($T_N = 245$ K).

{\SION} shows similar temperature dependence to {\SIOS} in single magnons and spin-orbit excitons, except that the sharp excitonic QP mode seen in {\SIOS} is absent in {\SION}. Although the single magnons are damped at high temperature, the substantial spectra weight left imply that the damped paramagnons remains at high temperature, which is consistent with previous reports.

\end{document}